\documentclass{nature}
\usepackage{lipsum}
\usepackage{amsmath}
\usepackage{amssymb}
\usepackage[normalem]{ulem}
\usepackage{lineno}
\usepackage{longtable}
\usepackage{booktabs}
\usepackage{svg}
\usepackage{adjustbox}
\usepackage{ragged2e}
\usepackage{booktabs}
\usepackage{multirow}
\usepackage{tabularx}
\usepackage{pdfpages}
\usepackage{caption}
\usepackage{subcaption}

\usepackage{pdflscape}
\usepackage{makecell}
\PassOptionsToPackage{hyphens}{url}\usepackage{hyperref}
\usepackage[vmargin=0.5in]{geometry}

\newcommand\ExtraSep
 
\renewcommand{\baselinestretch}{1}
\usepackage{caption}
\usepackage{graphicx}
\makeatletter
\let\saved@includegraphics\includegraphics
\AtBeginDocument{\let\includegraphics\saved@includegraphics}
\makeatother

\title{Healthcare Utilization Patterns Among Migrant Populations: Increased Readmissions Suggest Poorer Access. A Population-Wide Retrospective Cohort Study.}

\author{Elma Dervić$^{1,2,3}$, Ola Ali$^{2}$, Carola Deischinger$^{4}$,  Rafael Prieto-Curiel$^{2}$, Rainer Stütz$^{2}$, Ellenor Mittendorfer-Rutz$^{5,1}$, Peter Klimek$^{3,2,1}$}

\date{\today} 

\begin{document}

\maketitle

\begin{affiliations} 
\item Medical University of Vienna, Center for Medical Data Science, Institute of the Science of Complex Systems, Spitalgasse 23, 1090 Vienna, Austria;
\item Complexity Science Hub, Josefst\"adter Stra\ss e 39, 1080 Vienna, Austria;
\item Supply Chain Intelligence Institute Austria (ASCII), Josefstädter Straße 39, 1080 Vienna, Austria;
\item Medical University of Vienna, Department of Internal Medicine III, Clinical Division of Endocrinology and Metabolism,  Spitalgasse 23, A-1090 Vienna, Austria;
\item Department of Clinical Neuroscience, Division of Insurance
Medicine, Karolinska Institutet, Stockholm, Sweden;

\end{affiliations}

  $^*$Correspondence: dervic@csh.ac.at \\

\begin{abstract}

Equal access to health ensures that all citizens, regardless of socio-economic status, can achieve optimal health, leading to a more productive, equitable, and resilient society. Yet, migrant populations were frequently observed to have lower access to health. The reasons for this are not entirely clear and may include language barriers, a lack of knowledge of the healthcare system, and selective migration (a "healthy migrant" effect). We use extensive medical claims data from Austria (13 million hospital stays of approximately 4 million individuals) to compare the healthcare utilization patterns between Austrians and non-Austrians. We looked at the differences in primary diagnoses and hospital sections of initial hospital admission across different nationalities. We hypothesize that cohorts experiencing the healthy migrant effect show lower readmission rates after hospitalization compared to migrant populations that are in poorer health but show lower hospitalization rates due to barriers in access. We indeed find that all nationalities showed lower hospitalization rates than Austrians, except for Germans, who exhibit a similar healthcare usage to Austrians. Although around 20\% of the population has a migration background, non-Austrian citizens account for only 9.4\% of the hospital patients and 9.79\% of hospital nights. However, results for readmission rates are much more divergent. Nationalities like Hungary, Romania, and Turkey (females) show decreased readmission rates in line with the healthy migrant effect. Patients from Russia, Serbia, and Turkey (males) show increased readmissions, suggesting that their lower hospitalization rates are more likely due to access barriers.
Considering the surge in international migration, our findings shed light on healthcare access and usage behaviours across patients with different nationalities, offering new insights and perspectives. 
\newline
\end{abstract}

Migration $|$ Healthcare $|$ In-hospital data $|$ Hospitalisation Rate 
\maketitle


\section*{Introduction}

{
International migrants are nearly 300 million people worldwide, corresponding to approximately 3.6\% of the world's population \cite{UNMigration}. With an increasing international population living in a different country, certain topics are gaining relevance, particularly their integration into their hosting community \cite{OECDMigration, niva2023world}. Their integration fosters social cohesion and maximises the contributions of diverse populations to society, enhancing prosperity for all. Successful integration facilitates the smooth adaptation of newcomers to their host country. The integration of migrants involves many social dimensions, including labour, education and health \cite{OECDMigration, backman2018career, sirbu2020human, ortensi2022even}. Depending on the departure and arrival conditions, migrants might easily integrate into their new country or may struggle to access the labour market, could be left out of the education system or might not know how to receive medical care \cite{WHOReportHealthRefugeeMigrants}. 
}

{
Of critical relevance is that migrants tend to have lower access to health services. For example, data collected in 84 countries between 2018 and 2021 show that in half of the assessed countries, all migrants have access to all government-funded health services under the same conditions as nationals, regardless of their migratory status. In one-third of the countries, access to health services depends on migratory status; in 8\% of countries, migrants have access only to emergency health care services; and in 5\% of countries, they have no access to any government-funded health services \cite{WHOReportHealthRefugeeMigrants}. Migrants often have unaddressed healthcare needs, particularly in mental and dental health \cite{lebano2020migrants}. It is estimated that migrants in Europe have a high rate of unmet healthcare needs, although there are significant regional disparities \cite{kullamaa2023socio}. Even more challenging conditions related to access to healthcare services are experienced by refugee populations. Refugees rate their health lower than the resident population, predominantly female and Afghan refugees \cite{kohlenberger2019barriers}. For example, despite having high satisfaction with the Austrian health system, 20\% of male and 40\% of female refugees report unmet health needs \cite{kohlenberger2019barriers}. Health inequities of migrants are frequently reinforced by racism, anti-migration bias, healthcare-related prejudice, and unequal treatments \cite{pattillo2023experience}. 
}

{
The lack of a healthcare system for the migrant population is critical for many reasons. First, it has profound costs in terms of quality of life since some diseases become progressively severe, for example, myocardial infarctions or chronic diseases such as chronic obstructive pulmonary disease \cite{lewis2014impact, blinderman2009symptom}. Instead of receiving early attention, the burden of the disease, both in terms of health and resources, deteriorates. Second, with less access to healthcare services, patients also have less preventive healthcare, such as early treatment of obesity, vaccinations, etc. Some communicable diseases, such as Influenza, Chickenpox, Hepatitis B and C and sexually transmitted diseases, including syphilis, gonorrhoea, chlamydia, trichomoniasis and HIV, can have a more detrimental impact on communities with less access to the healthcare system and lower health literacy \cite{quinn2014health}. However, the burden of the lack of access to health services goes beyond the migrant population. In the United States, for example, vaccination rates for refugees and migrants are lower than for those born in the country \cite{WHOReportHealthRefugeeMigrants}. Refugees and migrants from various eastern African countries living in Australia indicated that they had limited access to immunisations, mostly because of language barriers \cite{WHOReportHealthRefugeeMigrants}. Similarly, most Syrian refugee children are not fully vaccinated \cite{WHOReportHealthRefugeeMigrants}. Beyond vaccination, there are also issues related to infectious diseases. Comprehensive, high-volume data concerning the health status of migrants and refugees increases the preparedness of the health system against mass infections and epidemic contagion \cite{matsangos2022health}. Finally, besides infectious diseases, there is also an issue with mental health. The lack of a correct diagnosis and mental health treatment could affect those who are in contact. For instance, nearly 100,000 individuals with a migration background in Germany have been diagnosed with dementia \cite{monsees2022dementia}.  
}

{
There are many reasons why migrants often have lower access to health services. For example, even if migrants are legally entitled to certain health services, they often face hidden costs, such as transportation or hiring translators, which reduce their chances of seeking care \cite{WHOReportHealthRefugeeMigrants}. However, the lack of access to health services goes beyond financial costs. For example, in Colombia, health-related expenditures among migrants and non-migrants living with HIV (and in contact with a medical facility) were lower for migrants, with the average per capita expenditure being US\$ 859 for migrants compared to US\$ 1796 for non-migrants, when adjusted for age and clinical characteristics \cite{WHOReportHealthRefugeeMigrants}. However, the challenge is not only based on offering health services to migrant populations or its cost. Migrants often suffer healthcare discrimination, language barriers and cultural disparities \cite{gil2021access}. Further, the rapid ageing of the world's population brings new challenges, especially for healthcare systems \cite{world2015age}. This demographic shift, characterised by an increasing proportion of elderly individuals, impacts the dynamics of health service demands and resource allocation. Furthermore, longer lifespans are accompanied by a shift in disease burden and a rise in healthcare and long-term care expenses \cite{euage}. An increasing migrant population, combined with an ageing country, requires a fundamental modification of healthcare institutions, practices, and cultural norms to meet these challenges promptly and effectively. 
}

{
The "healthy migrant effect" suggests that migrants might be a particularly resilient population \cite{ichou2019healthy, ma2020internal, holz2022health, kennedy2015healthy}. For example, the healthy migrant effect is frequently observed with lower mortality rates among migrants \cite{ichou2019healthy, ng2011healthy, razum1998low, dupre2012survival, holz2022health}. Turkish residents in Germany, for example, have persistently lower overall mortality rates, which extend into the second generation \cite{razum1998low}. However, these studies attribute this to the positive health selection of migrants, meaning healthier individuals are more likely to migrate, potentially beneficial cultural practices, and social support networks. Furthermore, the ``salmon bias'' hypothesis suggests that the observed lower mortality rates among migrant populations may be partly due to selective return migration. This bias describes older migrants moving back to their birthplace (similar to salmon fish returning to the freshwater streams where they were born to spawn) \cite{dunlavy2022investigating, ma2020internal, holz2022health}.
}

{
Observing the links between migration and access to healthcare is quite challenging for many reasons. Firstly, migration is highly dynamic, with repeat and return patterns observed frequently. Some people tend to move and return between different locations frequently, thus spending only a limited amount of time in any location \cite{PrietoColombiaMobility}. Recognizing these variations in migration patterns is crucial for understanding how migrants interact with the healthcare system. For instance, individuals who spend extended periods in a host country are more likely to establish regular healthcare-seeking behaviours and utilize a broader range of healthcare services compared to those with shorter stays. Therefore, accurately assessing healthcare utilization among migrants requires accounting for the diverse staying time in a host country and considering individual differences based on factors like gender, age, and country of origin. Secondly, there are many differences between the migrant and the residence population, including age structure, fertility rates and others. And finally, both migration and healthcare information are very sensitive data that are difficult to manipulate at large scales. 

}

{ 
The lack of healthcare is particularly critical for Austria. One in five persons living in Austria is a migrant (residents in Austria without Austrian citizenship), with an even higher ratio in larger cities like Vienna \cite{prieto2023diaspora}. In this study, migrants are defined by foreign nationality, a criterion unaffected by naturalization rates (which in Austria remained below 1\% from 2015 to 2019) \cite{statistikat2022}. We explored how nationality influences hospitalization and readmission rates among patients within the Austrian healthcare system. We hypothesized that significant disparities and patterns in inpatient healthcare access and utilization vary across different national backgrounds and that some nationalities have limited access to the healthcare system, worsening their quality of life. In particular, we hypothesize that readmission rates might uncover one of two reasons for decreased healthcare utilization: the healthy migrant effect and cultural barriers. Cohorts with the healthy migrant effect will exhibit lower admission and readmission rates post-hospitalization. In contrast, migrant populations with poorer health may have lower hospitalization rates due to access barriers, leading to higher readmission rates after initial hospitalization.
}

{
We analysed data from an electronic health registry covering almost 4 million residents of Austria, documenting over 12 million in-hospital stays spanning five years from 2015 to 2019 to test these hypotheses. A unique aspect of this dataset is the inclusion of patients' nationality information, enabling us to differentiate and compare the inpatient healthcare utilization patterns between Austrians and non-Austrians. Furthermore, this allowed us to explore the healthcare system usage patterns of migrant patients in Austria, offering valuable insights into the healthcare dynamics within diverse population segments. We introduced the \textbf{hospitalisation rate} to quantify how often different people use inpatient care, and the \textbf{readmission rate}, defined as the probability of patients returning to the hospital within a given time window following their initial visit \cite{milne1990can}. Hospital readmission rates can be used as an indicator of the quality of care \cite{pugh2021evidence, Kristensen2015Readmission}. 
For example, using the Nationwide Readmissions Database, it was found that increased social needs correlate with higher readmission rates \cite{bensken2021health}. Here, we account for individual differences in the duration of migrants' stays by introducing their Years of Life (YoL) in Austria, where YoL equals one for an entire year spent in the country and decreases proportionally with less time spent. Even though, in Austria, over 99\% of legal residents have health insurance coverage \cite{vintila2020migration}, we observed that migrants, particularly from remote countries, are less likely to gain access to health services.
}

\section*{Results}

\subsubsection*{Matching population groups}

{
We found substantial differences in hospitalisation rates (reflecting inpatient care utilization) and readmission rates (measuring how often patients are readmitted to the hospital within the next year) across cohorts of patients with different nationalities. In Austria, where nearly 20\% of the population are migrants, patients whose nationality is not Austrian comprise less than 10\% of all hospital patients and less than 10\% of the number of hospital nights. To consider varying age and sex- and gender-related structures, we matched different population cohorts (see the Methods section). Female patients of Austrian nationality have a hospitalisation rate of 0.22, and males have a rate of 0.23.
}

{
Female Syrians have a 1.5-fold (0.33) increased hospitalisation rate, whereas, for Russian females, the rate decreases by 50\% (0.12). However, hospital rates not related to pregnancy are significantly different for some nationalities. For instance, for female Syrians, it is 0.21. The highest hospitalisation rate among males is for patients from Austria, and the lowest hospitalisation rate among males is for patients from North Macedonia, which is 50\% smaller compared to Austrians, with a hospitalisation rate of 0.12. Looking at the top ten countries of origin with the most extensive diaspora in Austria, we could see that for German patients, females have a rate of 0.25, and males have a rate of 0.21, so they have a similar hospitalisation rate to their Austrian counterparts.

\begin{figure*}[!h]
\centering
{\includegraphics[width=0.99\linewidth]{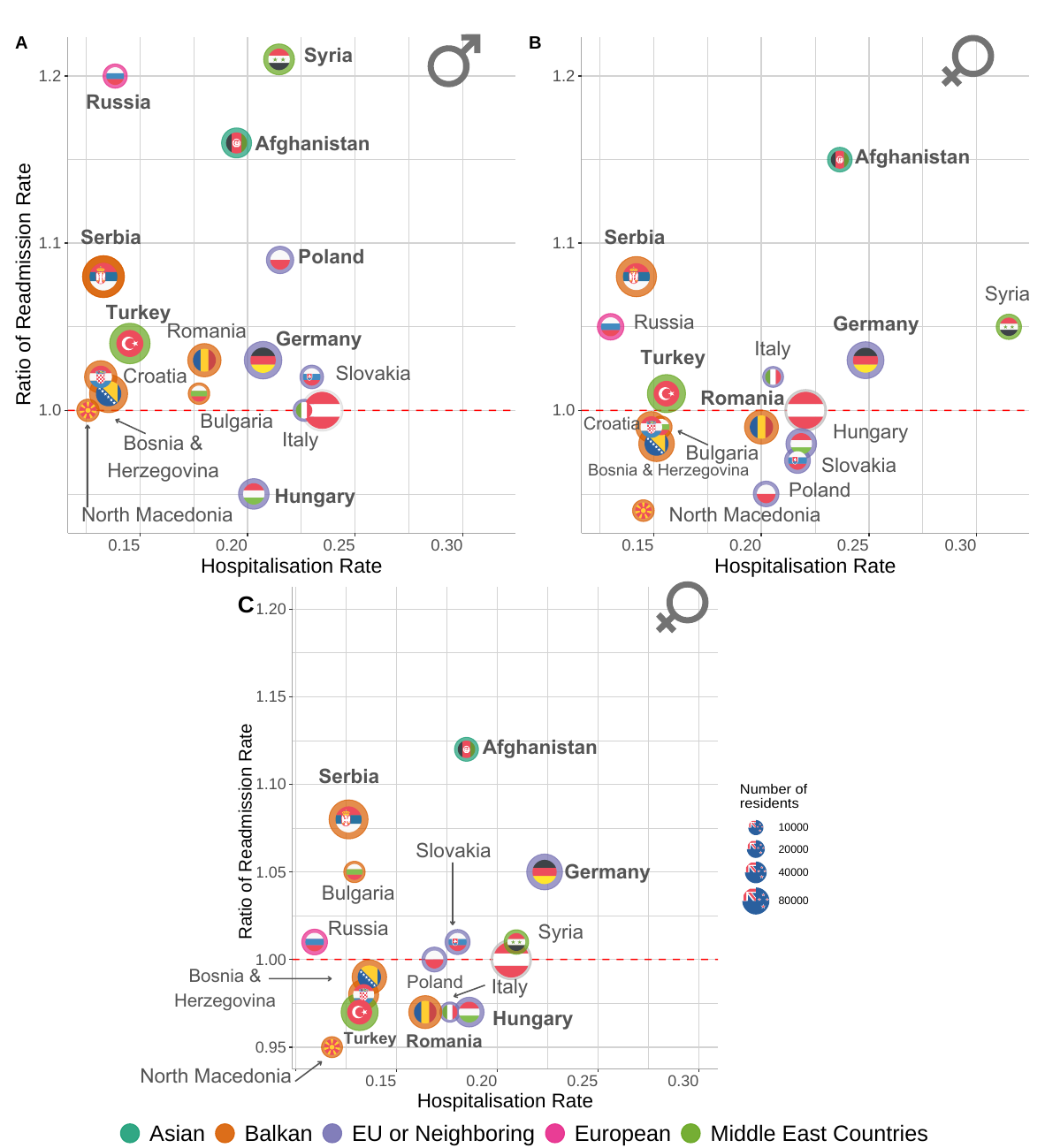}}

\caption{Hospitalisation rate of cohorts with different countries of origin and ratio of readmission rate in the next year of each cohort and their (age, sex, region of residence, and month of hospital visit) matched cohort of Austrian patients a) males, b) females, c) females - hospitalisation rate without pregnancy-related hospital stays. Countries in bold exhibit statistically significantly different readmission rates compared to matched Austrian counterparts. The size of the flags is proportional to the number of residents in Austria of each country of origin. The circle's colour around the flags depicts an area of the country (with respect to Austria).} 
\label{fig:HR_Final}
\end{figure*}
}

{
A crucial element observed is that we found similar hospitalisation and readmission rates for countries from the same area. For example, most Balkan countries, like Serbia, Croatia, Bosnia and Herzegovina, and North Macedonia, have comparable hospitalisation rates, Figure \ref{fig:HR_Final}. Thus, hospitalisation and readmission rates capture a collective behaviour observed for different groups of migrants in a hosting country.
}

{
Excluding pregnancy-related hospital stays for female patients, we found
that the highest hospital rate was for German patients (0.22), while the lowest was still for patients from Russia (0.11), Figure \ref{fig:HR_Final}c. The hospitalisation rate for Austrian female patients is 0.21. Countries like Bosnia and Herzegovina, Germany, Croatia, Serbia, Italy and Hungary had similar estimates for hospitalisation rates (including and excluding pregnancy-related hospital stays), where the difference was less than 15\%. The biggest difference between these two hospitalisation rates we identified for females from Syria, where 38\% of all hospital patients go to a hospital because of a pregnancy. However, in the case of Austria, only 9.5\% of female patients had only pregnancy-related hospital stays. 

}

{
Comparing baseline characteristics across the top ten countries, based on the number of patients, we found that almost in all cases, non-Austrian patients had a smaller number of diagnoses, fewer hospital stays and fewer hospital days in comparison with their matched Austrian peers, considering age, sex, the month of visit, region of residence (Table \ref{matched} and SI Figure 23).

\begingroup
\footnotesize
\renewcommand{\arraystretch}{1} 
\setlength{\tabcolsep}{3pt} 
 \begin{longtable}{cll|rr|rr|rr|rrrr}
\centering
Country & ISO2 & Sex & \multicolumn{2}{c}{\makecell{Number of \\Diagnoses}} & \multicolumn{2}{c}{\makecell{Number of \\Hospital Stays}} & \multicolumn{2}{c|}{\makecell{Number of \\ Hospital Days}} & \makecell{Total\\Number of\\Hospital\\Stays} & \makecell{RP} &   \makecell[c]{Ratio \\ of RP}\\ 
  \hline
 & & &  & Ratio & & Ratio & & Ratio & & &
\\
 \hline
  \hline
 
\multirow{2}{*}{Afghanistan} & \multirow{2}{*}{AF} & Male & 2.09 & $ 0.85^{***}$ & 3.17 & $ 0.67^{***}$ & 2.09 & $ 0.85^{***}$ & 9,802 & 0.36 & 1.16 \\ 
&  & Female & 2.57 & $ 0.94^{**}$ & 4.08 & $ 0.87^{***}$ & 2.57 & $ 0.94^{**}$ & 9,410 & 0.37 & 1.15 \\ 
  \multirow{2}{*}{\makecell{Bosnia \& \\ Herzegovina}} & \multirow{2}{*}{BA} & Male & 3.05 & $ 0.84^{***}$ & 6.52 & $ 0.91^{***}$ & 3.05 & $ 0.84^{***}$ & 24,322 & 0.47 & 1.01 \\ 
  & & Female & 2.66 & $ 0.81^{***}$ & 5.41 & $ 0.83^{***}$ & 2.66 & $ 0.81^{***}$ & 26,632 & 0.41 & 0.98 \\ 
  \multirow{2}{*}{Germany} & \multirow{2}{*}{DE} & Male & 3.04 & $ 0.82^{***}$ & 6.39 & $ 0.89^{***}$ & 3.04 & $ 0.82^{***}$ & 33,186 & 0.49 & 1.03 \\ 
    & & Female & 2.70 & $ 0.81^{***}$ & 5.37 & $ 0.84^{***}$ & 2.70 & $ 0.81^{***}$ & 39,154 & 0.43 & 1.03 \\ 
   \multirow{2}{*}{Croatia} & \multirow{2}{*}{HR} & Male & 2.92 & $ 0.81^{***}$ & 5.83 & $ 0.81^{***}$ & 2.92 & $ 0.81^{***}$ & 15,544 & 0.47 & 1.02 \\ 
  & & Female & 2.56 & $ 0.79^{***}$ & 5.22 & $ 0.81^{***}$ & 2.56 & $ 0.79^{***}$ & 16,463 & 0.41 & 0.99 \\ 
  \multirow{2}{*}{Poland} & \multirow{2}{*}{PL} & Male & 2.83 & $ 0.87^{***}$ & 7.22 & $ 1.06 $ & 2.83 & $ 0.87^{***}$ & 12,003 & 0.46 & 1.09 \\ 
   & & Female & 2.29 & $ 0.76^{***}$ & 5.63 & $ 0.93 $ & 2.29 & $ 0.76^{***}$ & 13,535 & 0.36 & 0.95 \\ 
   \multirow{2}{*}{Romania} & \multirow{2}{*}{RO}  & Male & 2.23 & $ 0.77^{***}$ & 4.52 & $ 0.83^{***}$ & 2.23 & $ 0.77^{***}$ & 11,600 & 0.38 & 1.03 \\ 
  &  & Female & 2.31 & $ 0.8^{***}$ & 3.65 & $ 0.7^{***}$ & 2.31 & $ 0.8^{***}$ & 23,676 & 0.35 & 0.99 \\ 
   \multirow{2}{*}{Serbia} & \multirow{2}{*}{RS} & Male & 3.41 & $ 0.93 $ & 7.53 & $ 0.95 $ & 3.41 & $ 0.93 $ & 33,162 & 0.52 & 1.08 \\ 
    &  & Female & 3.14 & $ 0.93^*$ & 6.43 & $ 0.89^{**}$ & 3.14 & $ 0.93^*$ & 40,161 & 0.47 & 1.08 \\ 
   \multirow{2}{*}{Slovakia} & \multirow{2}{*}{SK} & Male & 2.11 & $ 0.74^{***}$ & 4.07 & $ 0.69^{***}$ & 2.11 & $ 0.74^{***}$ & 4,760 & 0.37 & 1.02 \\ 
  & & Female & 2.20 & $ 0.78^{***}$ & 3.81 & $ 0.7^{***}$ & 2.20 & $ 0.78^{***}$ & 11,973 & 0.35 & 0.97 \\ 
   \multirow{2}{*}{Turkey} & \multirow{2}{*}{TR} & Male & 3.00 & $ 0.91^{***}$ & 6.51 & $ 0.96 $ & 3.00 & $ 0.91^{***}$ & 33,025 & 0.45 & 1.04 \\ 
  & & Female & 2.80 & $ 0.92^{***}$ & 5.22 & $ 0.84^{***}$ & 2.80 & $ 0.92^{***}$ & 42,898 & 0.40 & 1.01 \\ 
  \multirow{2}{*}{Hungary} & \multirow{2}{*}{HU}  & Male & 2.11 & $ 0.73^{***}$ & 4.89 & $ 0.87^{**}$ & 2.11 & $ 0.73^{***}$ & 8,913 & 0.36 & 0.95 \\ 
  & & Female & 2.23 & $ 0.78^{***}$ & 3.98 & $ 0.76^{***}$ & 2.23 & $ 0.78^{***}$  & 15,470 & 0.34 & 0.98 \\ 
   \hline
   \hline
\caption{Baseline characters table of selected counties of origin and ratio to their (age, gender, month of visit, region of residence ) matched cohorts of Austrian patients, tested if the difference of values is significant ( $^{*} p<0.05$, $^{**} p<0.01$, $^{***} p<0.001$). The readmission probability (RP) (second to the last column) shows the probability of having another hospital visit in the following year, while the last column, Ratio, shows the ratio of this probability and the readmission probability of the matched cohort of Austrian patients. A comprehensive table including all countries of origin can be found in Supplementary Information, Table 3. }
\label{matched}
\end{longtable}
\endgroup
}

{
We compared the readmission rate of non-Austrian patients with matched Austrian peers and found that the biggest differences between rates are for patients from Afghanistan, for females and males. Patients from Turkey, Germany, and Serbia have higher readmission rates for females and males. While patients from Bosnia \& Herzegovina, Romania, Croatia, Poland, and Slovakia have different rates for males and females, males have readmission rates comparing their Austrian peers, and this effect is the opposite for female patients. Hungarian female and male patients showed lower readmission rates than matched Austrian patients. 
}

{
To determine whether our findings are influenced by certain characteristics of the population of origin (such as the ratio of female residents, age structure, population size, cohort size of patients, and the size of the diaspora), we analyzed the correlation between these factors and hospitalisation and readmission rates. We did not find any significant correlation that could bias our findings (SI Figures 31 and 32).
}

\subsubsection*{Prevalence of First and All Diagnoses of Austrians and non-Austrians}

{
The information on the specific hospital departments or units within a hospital where patients are initially admitted can tell us what kind of care was provided and the nature of patient needs (SI Figure 17). Internal Medicine is the most frequent hospital department where female and male patients are admitted for initial hospital stay (32.1\% of all males and 28.7\% of females). Paediatrics is the most frequent for both sexes, aged 0 to 19 years old. The initial hospital visit of female patients aged 20-39 is in the Obstetrics and Gynecology section. Section Surgery is the most frequent for males 20 to 49 years old and females 40 to 59 years old. Internal medicine is the most common hospital department for male patients aged more than 50 years old and female patients older than 60. 
}

{
We found substantial differences in prevalences from some hospital departments by analyzing the ratio of prevalences of patients with nationalities different from Austrian and the corresponding stratum of Austrian patients, Figure \ref{fig:HSRatio}. For instance, Other Specialty Areas and General and Vascular Surgery sections have lower prevalence in non-Austrian cohorts across all age groups. Meanwhile, internal medicine and cardiology, internal medicine and nephrology, internal medicine, haematology and oncology, and radiology are more often hospital departments of initial admission in non-Austrian patients. We found similar patterns when we analyzed this on a country level for countries like Turkey, Serbia, Bosnia, and Herzegovina (Figure \ref{fig:HSRatio}). However, this is less prominent in Germany, and differences with the Austrian population are minor than in other non-Austrian populations.

\begin{figure*}[!ht]
\centering
{\includegraphics[width=1\linewidth]{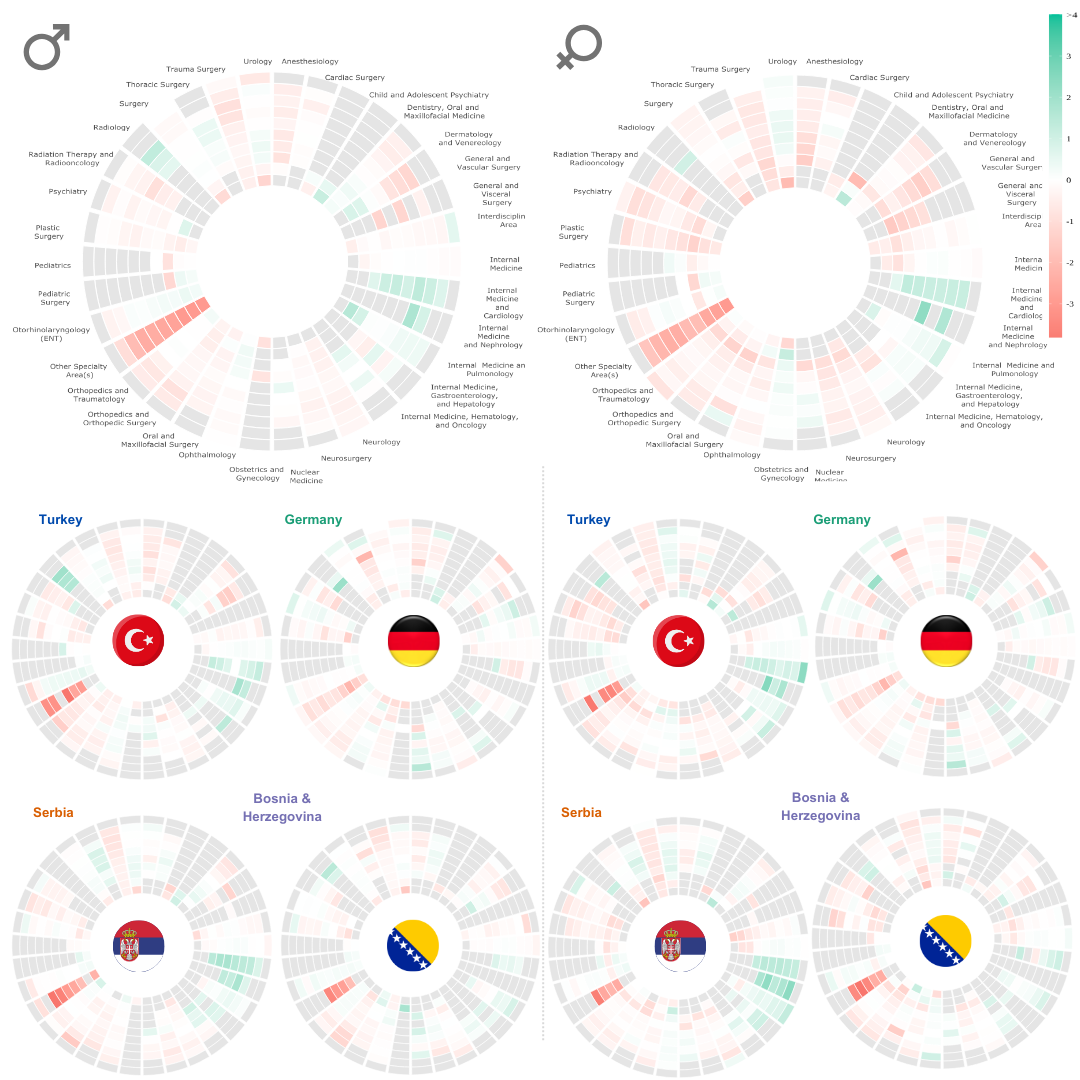}}
\caption{The top two graphs depict the ratio of prevalence of each hospital department of a first hospital visit in each 10-year age group (each circle) for non-Austrian nationality patients and Austrian nationality patients, males on the right side and females on the left (log scale). The bottom graphs show the results of 4 specific countries, males on the right side and females on the left (log scale). The grey colour indicates a stratum without patients in that hospital department. } 
\label{fig:HSRatio}
\end{figure*}
}

{
Primary and secondary diagnoses were classified using level-4 ICD10 codes (the global standard for medical coding), ensuring consistency across healthcare systems \cite{who}. The International Classification of Diseases (ICD) is a system produced by the World Health Organization that is crucial for tracking health trends and compiling international health statistics (details in the Methods section). To compare the most common reasons for initial hospital admissions, we calculated the prevalence of each ICD10 chapter of the primary diagnoses for each age group and sex for the Austrian and non-Austrian populations (SI Table 2 and Figure \ref{fig:FirstDiagChapter}). The primary diagnosis encodes the main reason for hospital admission. To compare these probabilities for different cohorts, we used the two-proportions z-test. The null hypothesis is that the corresponding probabilities for non-Austrian and Austrian patients are the same. We plotted the prevalence ratio of different ICD Chapters for non-Austrian and Austrian patients. However, we only plotted the ratio for cases where prevalences were significantly different, namely $p\_value < 0.05$ (the same results for all ICD10 codes except pregnancy-related diagnoses are shown in SI, Figure 14).

\begin{figure*}[!h]
\centering
{\includegraphics[width=1\linewidth]{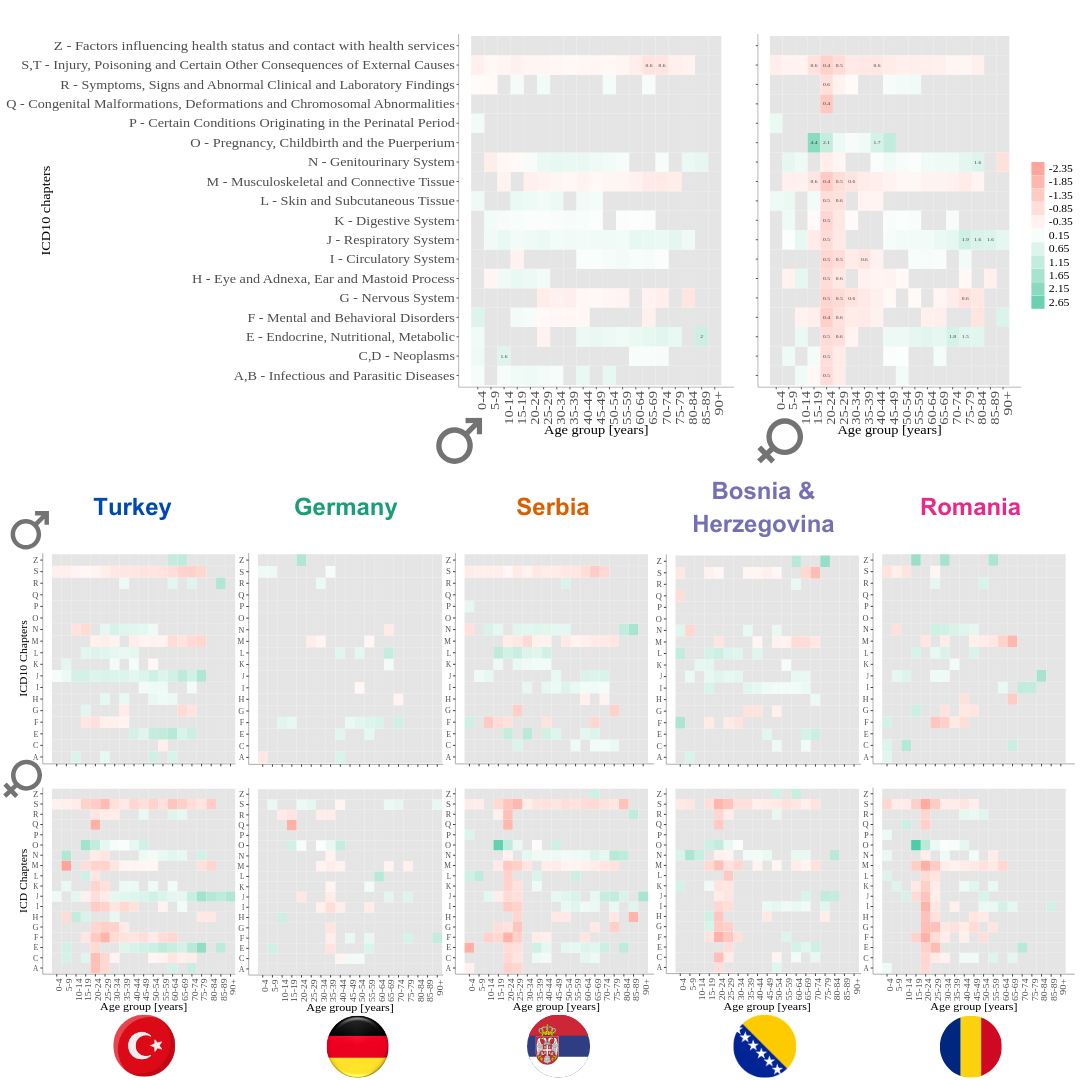}}

\caption{The top two graphs depict the ratio of prevalence of each type of diagnosis in each age group, non-Austrians/Austrians for the initial hospital stay, males on the right side, and females on the left (log scale). The bottom graphs show the results of four specific countries (the four largest groups of non-Austrian patients - Turkey, Germany, Serbia and Bosnia and Herzegovina), with males on the top and females on the bottom (log scale). The grey colour indicates a stratum without patients in that hospital department or that a certain probability of Non-Austrians is not significantly different from the corresponding probability of Austrians ($p-value >= 0.05$) } 
\label{fig:FirstDiagChapter}
\end{figure*}
}

{
The ratio between the prevalence of hospital diagnoses chapters between Austrian and non-Austrians ranges from 0.36 to 6.67. The most common ICD Chapter where non-Austrians have significantly smaller probability compared to Austrians are injury, poisoning and certain other consequences of external causes (S,T) and musculoskeletal and connective tissue (M). The respiratory system (J) and genitourinary system (N) were among the most common ICD Chapters, for which non-Austrians have a significantly higher probability than Austrians. Non-Austrian female patients in age groups 20 to 39 years old have a significantly smaller probability for all ICD Chapters, except for pregnancy, childbirth and the puerperium (ICD Chapter O). We found the same pattern for countries like Turkey, Serbia, Bosnia \& Herzegovina, and Romania. While this is not the case for female patients with German nationality, Figure \ref{fig:FirstDiagChapter}. 
}

{
Similar patterns to the diagnoses of the initial hospital admissions were found in the ratio between the prevalence of hospital diagnoses chapters between Austrian and non-Austrians for all hospital admissions \ref{fig:AllDiagChapter}. Concerning all hospital diagnoses, non-Austrian patients were more likely to be diagnosed with diseases of the respiratory (ICD chapter J) and genitourinary (N) system, next to endocrine, nutritional and metabolic diseases (E). However, diseases of the musculoskeletal system and connective tissue (M), injuries and poisonings (S,T) and mental and behavioural disorders (F) were used less often as primary hospital diagnoses in non-Austrians than in Austrians. 

\begin{figure*}[!h]
\centering
{\includegraphics[width=0.99\linewidth]{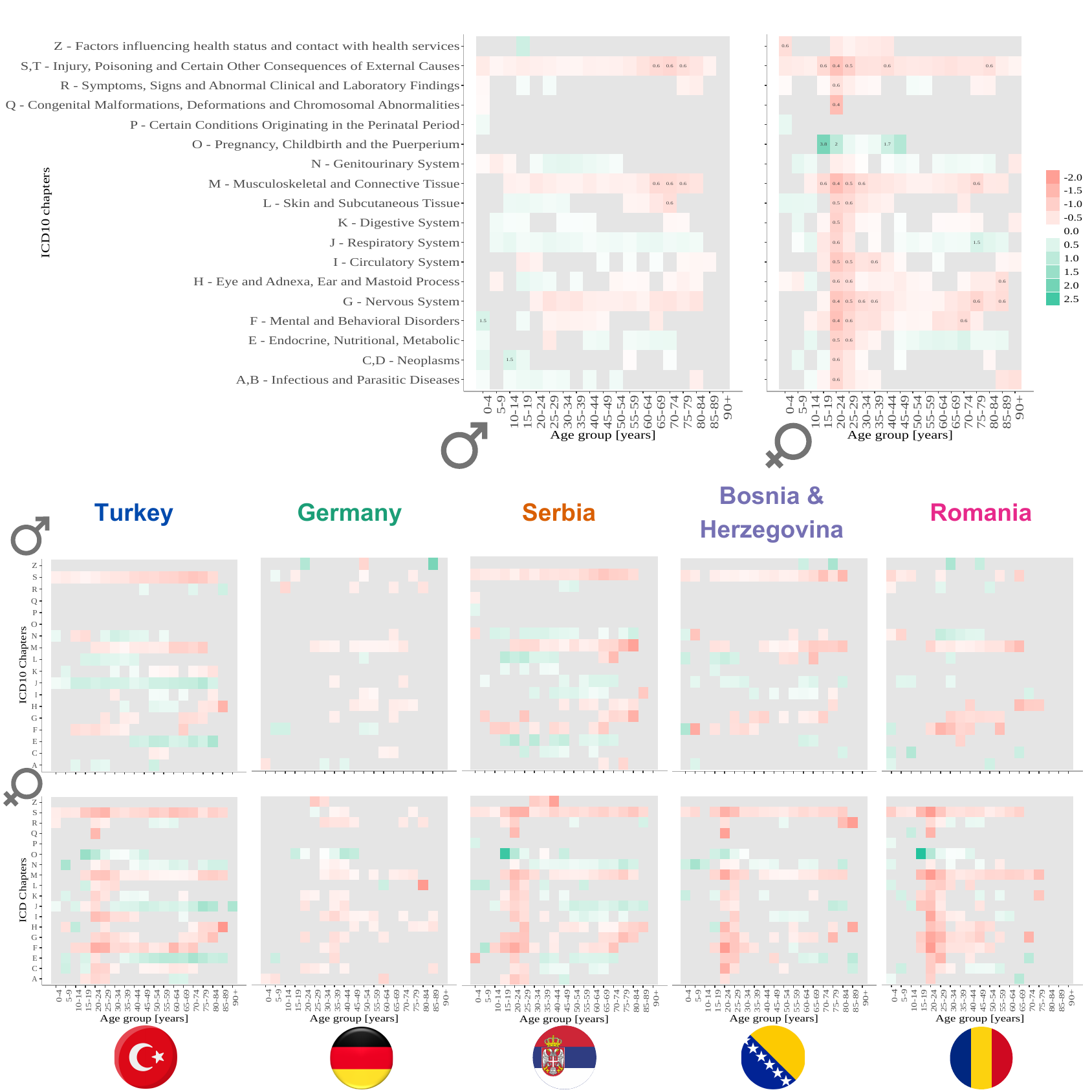}}

\caption{The top two graphs depict the ratio of prevalence of each type of diagnosis in each age group, non-Austrians/Austrians for all hospital stays, males on the right side, and females on the left (log scale). The bottom graphs show the results of four specific countries(the four largest groups of non-Austrian patients - Turkey, Germany, Serbia and Bosnia and Herzegovina), with males on the top and females on the bottom (log scale). The grey colour indicates a stratum without patients in that hospital department or that a certain probability of Non-Austrians is not significantly different from the corresponding probability of Austrians ($p-value >= 0.05$).  } 
\label{fig:AllDiagChapter}
\end{figure*}
}

\section*{Discussion}

{
In this extensive medical claims data analysis, we investigated differences in the usage of the hospital system by Austrians compared to non-Austrians. We focused on hospitalization rates and types of diagnoses to gauge how Austrians use the Austrian healthcare system compared to non-Austrians. Analyzing a comprehensive medical claims data set with information on patients' nationalities, we found that almost all nationalities, except for Germans, have lower hospitalization rates than Austrians. An exception is the hospitalization rate for non-Austrian women, which is higher than their Austrian counterparts. However, these hospitalizations are mostly comprised of admissions related to childbirth. We also saw clear differences in the diseases/diagnoses migrants and/or Austrians are admitted to the hospital. 
}

{
Three critical results for medical care were derived from the analyzed data. Almost all nationalities have lower hospitalization rates than Austrians (except for Germans living in Austria). In Austria, almost 20 \% of the population has a non-Austrian nationality but only comprises 9.4\% of the hospital patients and 9.79\% hospital nights \cite{prieto2023diaspora}. Although Germany is in second place regarding total patient count, it does lead in terms of the highest number of hospital stays. Male patients of all nationalities have lower hospitalization rates. On average, migrants go to the hospital less often (and potentially later in the development of a medical condition) as the readmission rate is higher in migrants than in Austrians \cite{milne1990can, bensken2021health}. At the same time, female patients from Germany, Afghanistan, and Syria have higher hospitalization rates in comparison to Austrians. However, in the case of hospitalization rates without pregnancy-related hospital stays, only German female patients have a higher hospitalization rate. Alarmingly, non-Austrian women between the ages of 20 and 39 years are much less likely to be admitted for any diagnosis other than those related to obstetrics (childbirth). This might indicate hesitance for non-Austrian women to use the medical system, which leads to worse medical care for female non-Austrians. 
 }
 
{
Our results show that almost all nationalities (except female German patients) have lower hospitalization rates. However, many nationalities also showed higher readmission rates, which may indicate that the 'healthy migrant' effect does not explain the lower hospitalization rate. Instead, a lower hospitalization rate may be present because of barriers to accessing the healthcare system. Hence, when migrants enter the healthcare system, there is a higher probability they will come again (higher readmission rate). This can also mean that hospital visits occur much later, leading to more severe cases and missed chances for proper and on-time prevention in the migrant population. Males from Syria, Russia, Afghanistan, Serbia, Poland, Turkey, and Germany have significantly higher readmission rates compared to matched Austrians. In females, the readmission rate differs greatly if we include or exclude pregnancy-related hospital stays. Namely, female patients with Turkish nationality have significantly higher readmission rates compared to Austrians if we include pregnancy. However, in the case of hospitalization without pregnancy-related stays, they have significantly lower readmission rates. At the same time, having a lower hospitalization rate and significantly lower readmission rate in female patients of Turkish nationality may indeed be driven by the "healthy migrant" effect. Female patients from Romania (and Hungary in case of stays without pregnancy-related hospitalizations) showed significantly lower readmission rates compared to Austrian females. In this case, the finding may be explained by the fact that these two countries are neighbouring countries, meaning these patients might also seek healthcare in their own country as EU citizens can access healthcare in other EU countries using their resident country's healthcare insurance \cite{eu_commission_2004}.
 }

{
Germans exhibit the most similar pattern in terms of admitted hospital department, first diagnosis, and all hospital diagnoses compared to Austrians. Except for Germans, we see clear differences in the hospital departments and diseases/diagnoses for which non-Austrians and Austrians are being admitted to the hospital. Non-Austrians of both sexes are more likely to be administrated in the Department of Radiology as well as all Departments of Internal Medicine. Female patients between the ages of 20 and 39 years frequent the Department of Obstetrics and Gynecology more often than their Austrian, age-matched counterparts. Moreover, there are some differences concerning the diagnoses for which the patients are admitted to the hospital. Young female non-Austrians are more often diagnosed with conditions related to Gynecology and Obstetrics, which is in line with birth rates being higher amongst non-Austrians \cite{statistikat2022}.
}

{
In non-Austrians of both sexes, the first diagnosis of a specific person was more likely to be from the chapters diseases of the genitourinary system (ICD chapter N), skin and subcutaneous tissue (L) and internal medicine (e.g. Diseases of the digestive System (K), diseases of the respiratory system (J), diseases of the circulatory system (I) and endocrine, nutritional and metabolic diseases (E). In terms of all coded hospital diagnoses, non-Austrian patients were more likely to be diagnosed with diseases of the respiratory system (J), diseases of the genitourinary system (N) and endocrine, nutritional and metabolic diseases (E). For instance, Turkish people of both sexes were more often admitted to the hospital due to respiratory issues. A slightly higher prevalence of smoking in Turkey (27.3\%) than in Austria (22.4\%) might partly explain the discrepancy \cite{eurostat}. 
}

{
Diseases of the musculoskeletal system and connective tissue (M), injury, poisoning and certain other consequences of external causes (S,T) and mental and behavioural disorders (F) were much less coded in non-Austrians than Austrians. Language and cultural barriers might hinder non-Austrians from seeking help for mental and behavioural disorders. It has been shown that difficulties in understanding and articulating emotional stress hinder help-seeking for mental conditions and that there is a clear association between a lack of language proficiency and underuse of psychiatric help \cite{ohtani2015language}. This is especially troubling for refugees, who are more likely to suffer from psychiatric diagnoses such as post-traumatic stress disorder, depression, anxiety, and panic disorder \cite{bhui2003traumatic, mollica2001longitudinal, sonderskov2021terrorism, caroppo2023migrants}.
}

{
This study's main strength lies in its extensive unique data. It comprehensively captures in-hospital data for approximately four million individuals. However, there are certain limitations to consider, primarily stemming from data quality and accessibility constraints. The data does not include information on outpatient medical services, prescriptions, or lifestyle factors. Furthermore, the absence of socioeconomic data for the individual patients means that the potential effects of socioeconomic factors remain unknown. Yet, results regarding hospital departments may be biased due to the different demographics of mostly Viennese (bigger) hospitals. Additionally, one more limitation of streaming data is that the database covers only five years and that we did not include patients with more than one nationality. We must emphasize that another potential limitation of this study lies in the estimates of ``years of life'' ($YoL$). We do not have data on the number of days Austrians spend in Austria, so we assumed that $YoL$ for all Austrians is 1. We do not have data on days spent for other nationalities for 2019, but only for 2023. This means that our estimates of $YoL$ for 2019 are based on data from 2023. Additionally, the flow estimates of the non-Austrian resident population assume a stable flow between 2023 and 2019, which is not valid in some cases, such as in Syria. 
}

{
Non-Austrians use the healthcare system less but are more likely to be readmitted, potentially due to more severe illness. Non-Austrians of both sexes are less likely to be diagnosed with an ICD10 diagnosis from chapter mental and behavioural disorders (F) and more often hospitalized with diseases from the ICD10 chapters of internal medicine, genitourinary diseases, and skin disorders. A prominent healthcare gap concerns young non-Austrian women, who are much less likely to go to a hospital than their Austrian peers. Similarly, results of the Refugee Health and Integration Survey (ReHIS) in Austria, published in 2019, indicated the refugee's self-rated health to be below the self-rated health of Austrians. Listed barriers were long waiting lists, lack of knowledge about doctors, and a language barrier \cite{kohlenberger2019barriers}. Less access to healthcare for non-Austrians is critical in terms of less preventative healthcare such as vaccinations, higher rates of communicable diseases or chronic diseases aggravating due to the lack of treatment. This might not only impact a person's physical health but also mental health and quality of life.
}

{
Early prevention plays a crucial role in maintaining an individual's well-being and public health. This is especially crucial considering that non-Austrians in our cohort are, on average, 13.9 years younger than Austrians. Prevention involves taking proactive measures to identify and mitigate potential health risks before they develop into more severe conditions. This approach is particularly significant in managing chronic diseases, where early intervention can dramatically reduce the severity and progression of the illness. Targeted research on chronic diseases can provide strategic, data-based solutions for preventing diseases, addressing some of the critical issues associated with this demographic evolution \cite{world2015age}. 
}

{
Taken together, the different patterns we found in hospitalization and readmission rates for people from countries other than Austria show that most of the migrant population using fewer healthcare systems cannot be driven only by controversial hypotheses of "healthy migrant" or only due to cultural barriers. Instead, utilization of the healthcare system in migrant populations has to be studied separately for different countries, as we appear to see different healthcare utilization patterns in diverse cohorts of migrants.
}

{
Future work should be dedicated to further investigating healthcare utilization and the comorbidities and multimorbidity patterns of patients of different nationalities. Future projects should develop strategies to reduce the barrier for non-Austrians to use the healthcare system, especially the outpatient system. Examples could be better translation services in hospitals and doctor's offices using web-based interpretation services, courses on how to navigate the Austrian healthcare system, particularly specialist care, or language courses to bridge the language gap.
}

\section*{Methods} 

\subsubsection*{In-hospital Data}

{
The analyzed dataset captures a comprehensive overview of hospital admissions in Austria, encompassing around 13 million cases from 4 million people ($N=3,999,832$) from January 2015 to December 2019. The demographic composition of this population is balanced in terms of gender (with 46\% male and 54\% female patients). It has records of each hospitalisation, including unique patient identifiers, sex and age (in a resolution of five years), along with primary and secondary diagnoses, the dates of admission and discharge, the nature of the discharge (such as standard release or transfer to another facility, etc.), region of residence, region of hospital \cite{haug2020high,dervic2021effect}, and nationality of the patients. While the primary diagnosis represents the main reason for hospitalisation, any additional diagnoses during the stay are recorded as secondary. 
}

{
Primary and secondary diagnoses in this dataset are classified using level-4 ICD10 codes, the global standard for medical coding and ensuring consistency across healthcare systems \cite{who}. The International Classification of Diseases (ICD) is a system produced by the World Health Organization that is crucial for tracking health trends and compiling international health statistics. This dataset has 12,040 distinct medical diagnoses, all classified under ICD codes. The five most frequently occurring primary diagnoses in ICD10 coding within this dataset are detailed in SI Figure 5. 
}

{
The mean age of the whole cohort is 52.8 SD 24.4 years. Yet, there are differences across nationalities. For Austrian patients, the mean age is 53.9 SD 24.4 years, and for non-Austrians, it is 40 SD 18 years, Figure \ref{fig:NumberOfHStays_Age}.

\begin{figure*}[!h]
\centering
{\includegraphics[width=0.8\linewidth]{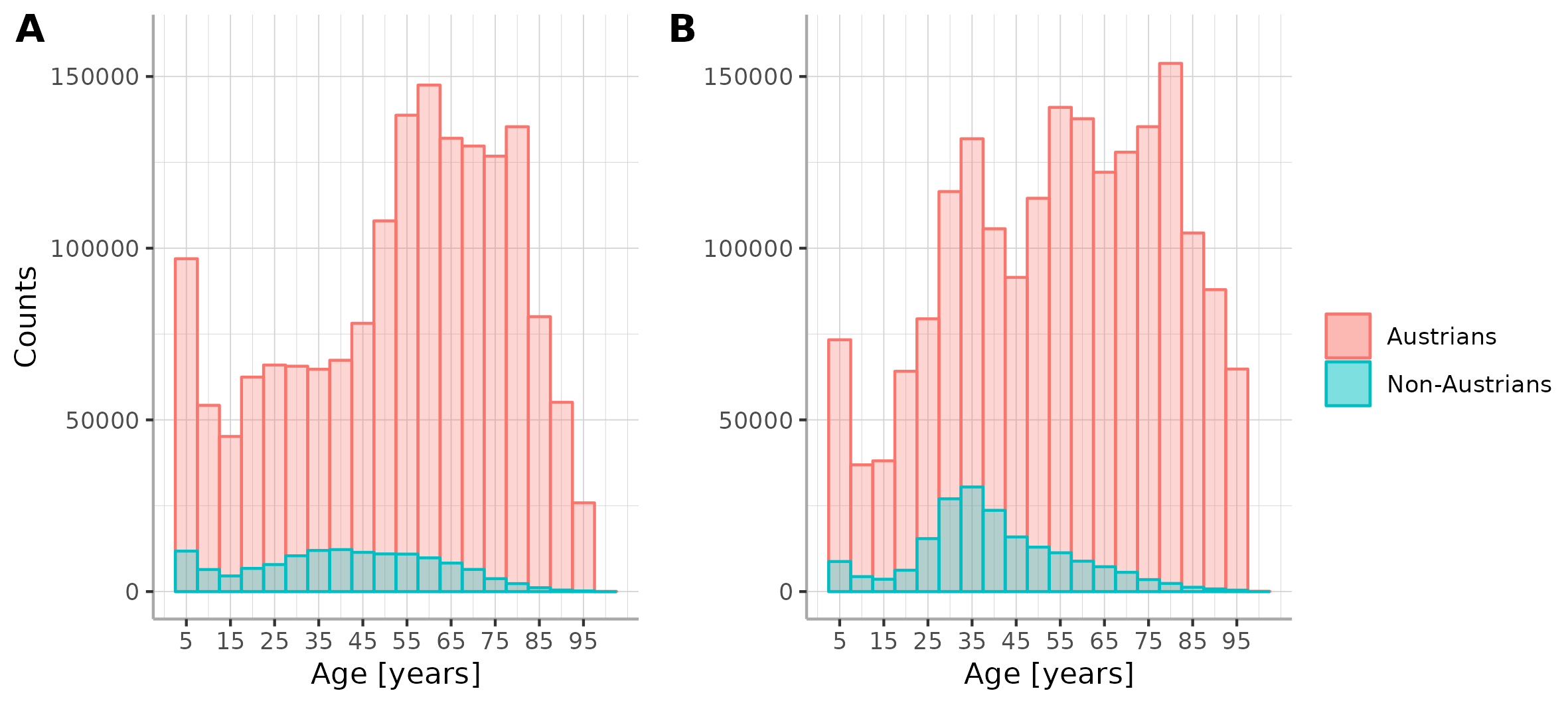}}
\caption{Age distribution a) males b) females} 
\label{fig:NumberOfHStays_Age}
\end{figure*}
}

{
We excluded all patients with more than one nationality from the analysis, which was 1.63\% of all patients. In the analysed cohort, patients whose nationality is not Austrian comprise 379,854 individuals, Figure \ref{fig:allpats}, accounting for 9.4\% of the entire cohort. Within this subset, females represent 57.7\%, while males constitute 42.3\%. The majority of these countries have a higher proportion of female patients compared to male patients. Specifically, there are 32\% more female patients of Turkish origin than their male peers, 26\% more from Germany, and 28\% more from Serbia. Notably, among the top ten countries of origin with the largest patient numbers, the cohort from Afghanistan is an exception, exhibiting 15\% fewer female patients compared to males of the same nationality. The female cohort from Afghanistan makes up around 35.8\% of the Afghan population, with around 19,795 females compared to 35,365 males. 

}

{

\begin{figure*}[!h]
\centering
{\includegraphics[width=0.95\linewidth]{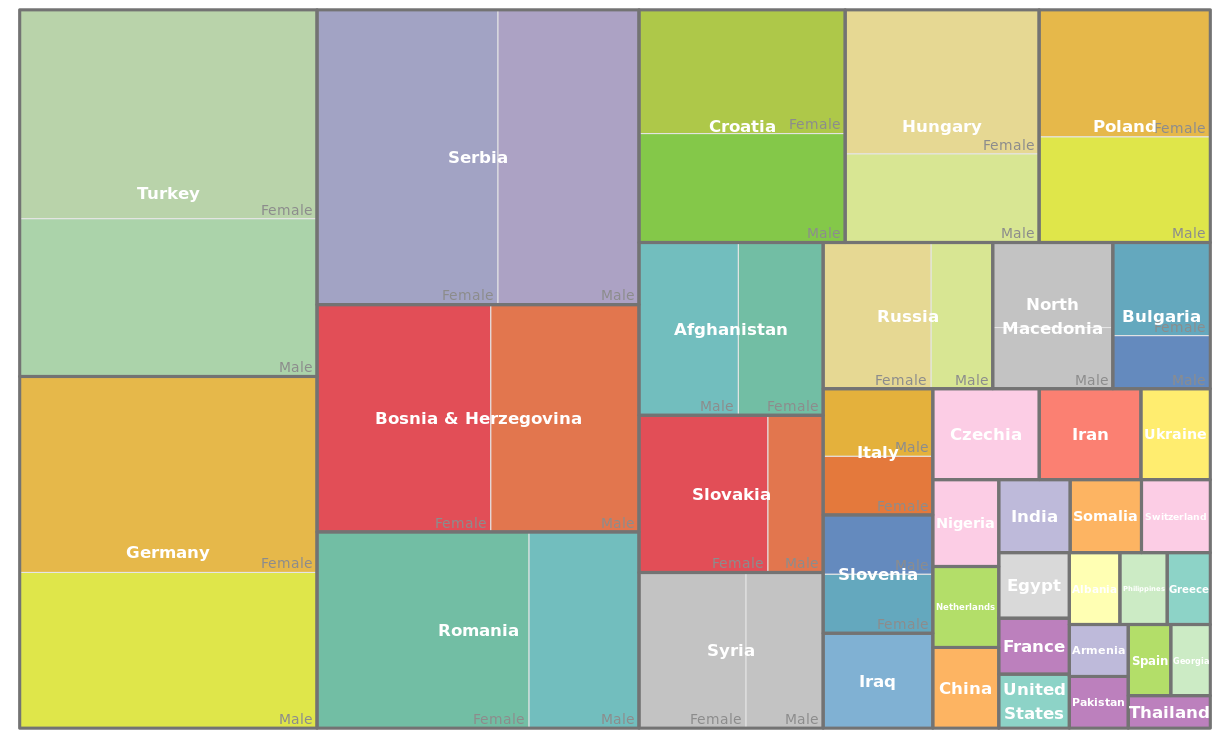}}
\caption{Proportion of each country of origin in migrant patients.} 
\label{fig:allpats}
\end{figure*}

\begingroup
\renewcommand\arraystretch{0.87} 
 \begin{longtable}{lc|c|c|c}
\centering
Country & ISO2 & Total Number & percentage & \makecell{Ratio of Number \\Females and Males}\\ 
 \hline
  \hline
Turkey & TR & 45,463 & 11.43 & 1.32 \\ 
 Germany & DE & 43,610 & 10.97 & 1.26 \\ 
 Serbia & RS & 39,580 & 9.95 & 1.28 \\ 
 Bosnia \& Herzegovina & BA & 30,514 & 7.67 & 1.17 \\ 
 Romania & RO & 26,355 & 6.63 & 1.93 \\ 
 Croatia & HR & 20,006 & 5.03 & 1.14 \\ 
 Hungary & HU & 18,816 & 4.73 & 1.63 \\ 
 Poland & PL & 16,613 & 4.18 & 1.20 \\ 
 Afghanistan & AF & 13,270 & 3.34 & 0.86 \\ 
 Slovakia & SK & 12,076 & 3.04 & 2.34 \\ 
  \hline
 Austria & AT & 3,669,104 &  & 1.15\\ 
  \hline
\caption{Total number of patients, percentage of the non-Austrian cohort, Ratio of number females and males for different countries of origin. Extended table with all countries available in Supplementary Information, Table 1.} 
\label{}
\end{longtable}
\endgroup

}

{
Non-Austrian patients had approximately 1.3 million stays, which is 9.79\% of all recorded stays, from which 46.31\% stays of males and 53.69\% of females, Figure \ref{fig:allpats}. While Germany ranks second in terms of patient count, it significantly leads as the country with the highest number of hospital stays, Figure \ref{fig:NumberOfPatStays}.

\begin{figure*}[!h]
\centering
{\includegraphics[width=0.99\linewidth]{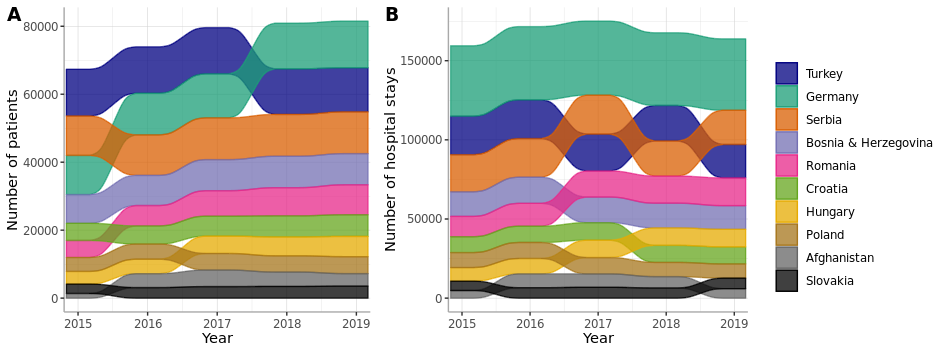}}
\caption{Number of a) patients and b) hospital stays across top ten counties of origin, determined by the patient counts from in-hospital data} 
\label{fig:NumberOfPatStays}
\end{figure*}
}

\subsubsection*{Estimating the years of life spent in the country}

{
We take into account that some migrants spend most of their time in the host country, but others arrive late (for example, students who frequently arrive during the summer break), and others tend to spend only part of their time in the host country. To account for these individual differences, we calculate the years of life spent by a person $YoL$ such that $YoL = 1$ if a person spends the entire year in Austria and has a smaller value with less time spent in the country. This measure is similar to the person-years measure used in migration studies, which accounts for the time spent by migrants when analyzing outcomes such as economic earnings and mortality \cite{constant2003self,wallace2015mortality}. The YoL of a non-Austrian resident depends on the country of origin, age and sex, so we model the YoL estimates as a function of those variables. We assume the YoL pattern is conserved per age group, sex, and country between 2023 and 2019 (details in the SI). We calculate the YoL spent by non-Austrian residents by using the YoL ratio across the country, sex, and age of 2023, and we apply it to the non-resident population in 2019 (details in the SI). 

Data on the number of days spent by residents in Austria per age, sex, and country of origin was provided by the Federal Ministry of Interior of Austria, Bundesministerium für Inneres (BMI) for 2023 \cite{prieto2023diaspora}. Among the 1,654,086 residents with non-Austrian nationality, 49\% are female and 51\% are male. The average age among these residents is 39.75 years, with females averaging 40.16 years and males 39.37 years, Figure \ref{fig:bmidata}. Data on the whole population of Austria and residents with non-Austrian nationality for 2015 to 2019 is provided by Statistik Austria via the online WEB application STATcube - Statistische Datenbank \cite{statsat}.

\begin{figure*}[!h]
\centering
{\includegraphics[width=1\linewidth]{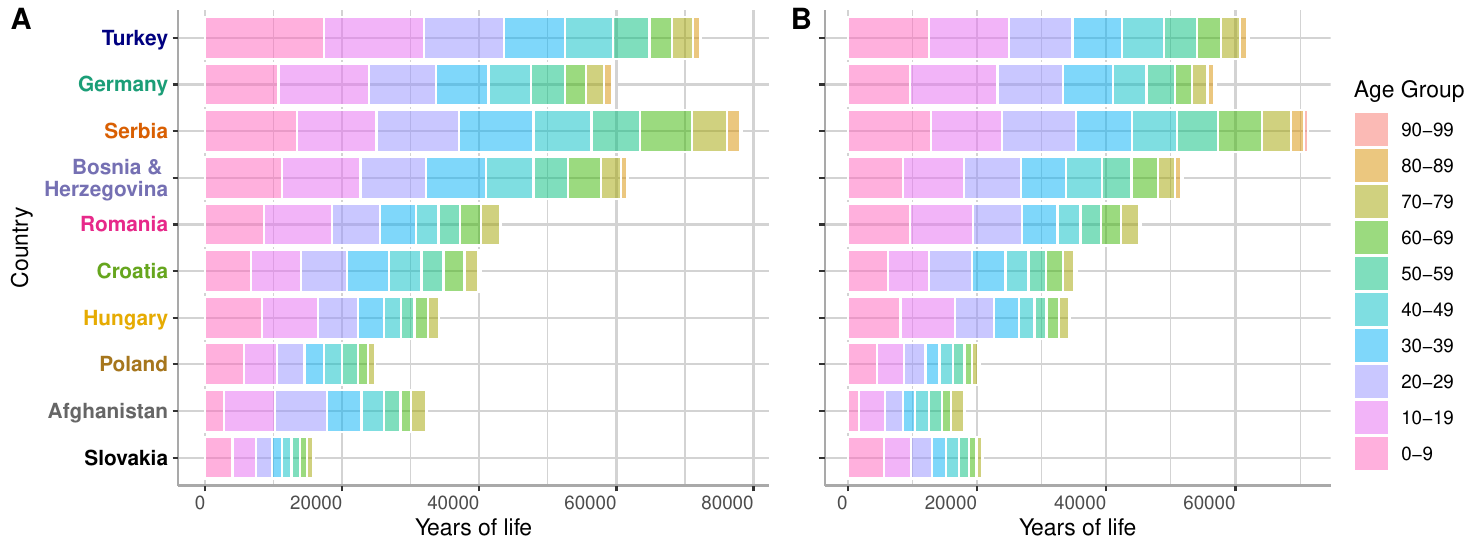}}
\caption{Estimated number of years of life from the top ten countries, determined by the patient counts from in-hospital data in 2019 in Austria a) males b) females } 
\label{fig:bmidata}
\end{figure*}
}

\subsubsection*{First hospital admissions}

{
We analyzed in-hospital data, concentrating specifically on patients' first hospital admissions, Figure \ref{fig:NumberOfHStays}. 

\begin{figure*}[!h]
\centering
{\includegraphics[width=0.6\linewidth]{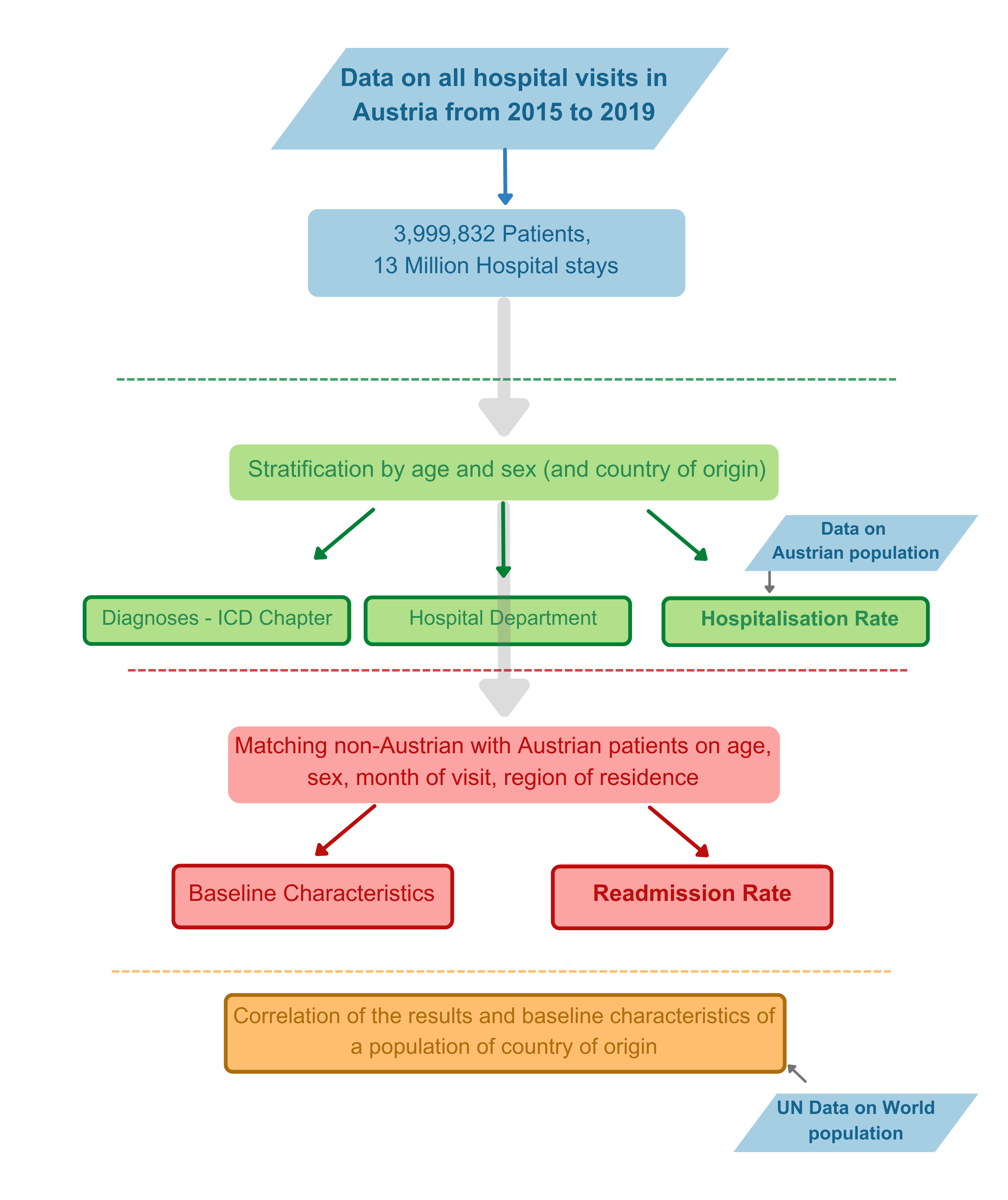}}
\caption{Workflow of the research presented in this article} 
\label{fig:NumberOfHStays}
\end{figure*}
}

{
We examined the primary diagnoses as the reason for hospital visits. Primary diagnoses in this dataset are coded using the level-4 ICD10 codes. We organized all ICD10 4-digit diagnosis codes into ICD Chapters (based on the first letter of each code), resulting in a total of 21 chapters. All ICD10 chapters are listed for better visibility in Supplementary Information Table 2. We then calculated the prevalence of each primary diagnosis chapter across each 10-year age group, sex, and nationality. Furthermore, to assess the differences in these probabilities across various cohorts, we employed the two-proportions z-test. This test's null hypothesis states that the respective probabilities for Austrian and non-Austrian patients are identical. The null hypothesis is rejected for cases where $p\_value < 0.05$. Subsequently, we calculated the ratio of these prevalences to the ones observed in Austrian patients for cases where we found a significant difference ($p\_value < 0.05$). This allowed us to draw comparisons between non-Austrian and Austrian patients in terms of the reason for their initial hospital admissions.
}

{
Moreover, the data shows 38 different hospital departments, such as Internal Medicine, Trauma Surgery, Dermatology, and Venereology (All sections are listed in SI, Table 3). We extended our analysis to include the prevalence of admissions in each hospital session, again stratifying the data by 10-year age groups, sex, and nationality. 
}

{
We introduced the \textbf{hospitalisation rate} to quantify how often different cohorts utilize (at least once) inpatient care. This rate is determined by dividing the number of hospitalized patients by years of life (approximation for the number of residents) in each specific age group, sex, and nationality category during 2019. We calculated the average hospitalisation rate for each nationality by computing a weighted average across age groups. The weights were assigned proportionally to the frequency of the general population within each age group. Additionally, for any group (stratum) with fewer than 50 patients, we replaced its hospitalisation rate with the corresponding rate for Austrian patients.
}

{
To accurately compare non-Austrian cohorts with their Austrian counterparts, we matched groups based on age, sex, the month of the hospital visit, and patient region in a ratio of 1:1. Following this matching process, we calculated the average number of hospital days, the number of hospital stays, and the number of diagnoses for each matched cohort during the observed period. Additionally, we tested if these estimates are significantly different for specific nationalities and its matched cohorts of Austrian patients, using independent samples T-Test (with a $p\_value < 0.05$).
}

{
The \textbf{readmission rate} is defined as the probability of patients returning to the hospital within a given time window following their initial visit. It is often used as a quality indicator in healthcare, with a high readmission rate potentially signalling issues in patient care. It can also reflect broader systemic problems, like lack of access to health care, which can slow a patient's recovery. Hence, a high readmission rate can indicate patients coming to the hospital too late with severe conditions. For our analysis, we calculated this rate for patients from different countries and their matched Austrian cohorts and for time windows ranging from three months to one year. To draw comparisons, we then computed the ratio of each country's readmission rate to the readmission rate of its corresponding matched Austrian cohort.
}

\subsubsection*{Ethics}

We made secondary use of a research database containing medical claims records, which is securely managed by the Federal Ministry of Health. It is important to note that measures have been implemented to guarantee the anonymity of individuals within this database. It is a consolidated research database accessible only to authorized partners who adhere to stringent data protection policies. Our use of this data is conducted in collaboration with the data provider and follows established agreements.

The data in this database do not include any personal identifiers, such as names, postal codes, or dates of birth. Additionally, all members of our research team have committed to maintaining confidentiality and complying with relevant data protection regulations through a signed agreement.

\section*{Acknowledgments}

\section*{Data Availability}
The raw and processed patient data are not available due to privacy laws. The dataset is safeguarded by the Austrian Federal Ministry of Health and made accessible to research institutions under strict data protection regulations. To gain access to this data, researchers have to find individual arrangements with the Austrian Federal Ministry of Health. 

 \section*{Code Availability}
Custom code for the analysis is available per request from the authors.

\section*{Competing Interests}
The authors declare no competing interests.

\section*{Author Contributions}
E.D., P.K., and R.P.C designed research; E.D. carried out the analysis, produced the plots and graphics, and drafted the manuscript; O.A. conducted analysis on years of life; E.D. and C.D. discussed the results;  E.D. P.K., R.P.C, O.A. C.D., R.S., E.M.R. analyzed the results. All authors wrote the manuscript.


\renewcommand{\baselinestretch}{1.5}


\renewcommand{\baselinestretch}{1.5}
\section*{References}




\section*{Figure Legends}

\textbf{Figure 1}: Hospitalisation rate of cohorts with different countries of origin and ratio of readmission rate in the next year of each cohort and their (age, sex, region of residence, and month of hospital visit) matched cohort of Austrian patients a) males, b) females, c) females - hospitalisation rate without pregnancy-related hospital stays. The size of the flags is proportional to the number of residents in Austria of each country of origin. The circle's colour around the flags depicts an area of the country (with respect to Austria).

\noindent\textbf{Figure 2}: The top two graphs depict the ratio of prevalence of each hospital department of a first hospital visit in each 10-year age group (each circle) for non-Austrian nationality patients and Austrian nationality patients, males on the right side and females on the left (log scale). The bottom graphs show the results of 4 specific countries,  males on the right side and females on the left (log scale). The grey colour indicates a stratum without patients in that hospital department. 

\noindent\textbf{Figure 3}: The top two graphs depict the ratio of prevalence of each type of diagnosis in each age group, non-Austrians/Austrians for the initial hospital stay, males on the right side, and females on the left (log scale). The bottom graphs show the results of four specific countries (the four largest groups of non-Austrian patients -  Turkey, Germany, Serbia and Bosnia and Herzegovina), with males on the top and females on the bottom (log scale). The grey colour indicates a stratum without patients in that hospital department or that a certain probability of Non-Austrians is not significantly different from the corresponding probability of Austrians ($p-value >= 0.05$)

\noindent\textbf{Figure 4}: The top two graphs depict the ratio of prevalence of each type of diagnosis in each age group, non-Austrians/Austrians for all hospital stays, males on the right side, and females on the left (log scale). The bottom graphs show the results of four specific countries(the four largest groups of non-Austrian patients -  Turkey, Germany, Serbia and Bosnia and Herzegovina), with males on the top and females on the bottom (log scale). The grey colour indicates a stratum without patients in that hospital department or that a certain probability of Non-Austrians is not significantly different from the corresponding probability of Austrians ($p-value >= 0.05$)

\noindent\textbf{Figure 5}: Age distribution a) males b) females

\noindent\textbf{Figure 6}: Proportion of each country of origin in patients with migration background

\noindent\textbf{Figure 7}: Number of a) patients and b) hospital stays across top ten counties of origin, determined by the patient count from in-hospital data

\noindent\textbf{Figure 8}: Estimated number of years of life from the top ten countries, determined by the patient counts from in-hospital data in 2019 in Austria a) males b) females 

\noindent\textbf{Figure 9}: Workflow of the research presented in this article


\section*{Table Legends}

 \noindent\textbf{Table 1}: Baseline characters table of selected counties of origin and ratio to their (age, gender, month of visit, region of residence ) matched cohorts of Austrian patients, tested if the difference of values is significant ( $^{*} p<0.05$, $^{**} p<0.01$, $^{***} p<0.001$). The readmission probability (RP) (second to the last column) shows the probability of having another hospital visit in the following year, while the last column, Ratio, shows the ratio of this probability and the readmission probability of the matched cohort of Austrian patients. A comprehensive table including all countries of origin can be found in Supplementary Information, Table 3.
 
 \noindent\textbf{Table 2}: Total number of patients, percentage of the non-Austrian cohort, Ratio of number females and males for different countries of origin. Extended table with all countries available in Supplementary Information, Table 1


\end{document}